# A Time-to-Digital Converter-based Correction Method for Charge Measurement through Area Integration

Jiajin Zhang, Lei Zhao, *Member, IEEE*, Ruoshi Dong, Peipei Deng, Cong Ma, Jiaming Lu, Shubin Liu, *Member, IEEE*, Qi An, *Member, IEEE*

*Abstract*—A high-precision charge measurement can be achieved by the area integration of a digitized quasi-Gaussian signal after the signal passes through the shaper and analog-to-digital converter (ADC). The charge measurement contains an error due to the uncertainty of the first sampled point of a signal waveform. To reduce the error, we employ a time-to-digital converter (TDC) to measure the uncertainty precisely, and we design correction algorithms to improve the resolution of the charge measurement. This work includes analysis and simulations of the proposed algorithms and implementation of them in an FPGA device. Besides, the tests are also conducted to evaluate the performance of the correction method. Test results indicate that the resolution of the charge measurement is successfully improved from 2.31‰ to 1.26‰ by using a signal from the shaping circuit (with the amplitude of 2 V, and leading and trailing edges of about 80 ns and 280 ns, respectively) digitized at the sampling rate of 62.5 Msps.

*Index Terms*—Charge measurement, correction algorithm, fine time, TDC

## I. INTRODUCTION

CHARGE measurement has always been an important task and a research hot spot in the nuclear physics field. In order to achieve a high-precision charge measurement, several techniques were proposed, including the waveform digitization technique based on switched capacitors arrays (SCA) [1-5], the time-over-threshold technique (TOT) [6-9], as well as the amplification, shaping and A/D conversion technique [10-18]. Among all these methods, the mainstream solution is to digitize the signal after amplification and shaping, and then, to detect the signal peak. For instance, in the Daya Bay Reactor neutrino experiment [11,13,14] and the large high altitude air shower observatory (LHAASO) water Cherenkov detector arrays (WCDA) [12,15-18], this method was employed to deal with a large dynamic-range PMT signal output (Daya Bay: 1-9000 photoelectrons & LHAASO: 1-4000 photoelectrons). Using the mentioned method, a high-resolution charge measurement can be achieved (e.g., LHAASO WCDA electronics: 1% rms @4000 photoelectrons) [16]; however, this method is still limited by a noise of the analog circuit and analog-to-digital converter (ADC), and a peak detection error. An alternative method is to obtain the charge information through the summation of the sampling points from an ADC to calculate a "waveform area" instead of the peak detection [19-23].

Still, there remains the considerable interest in improving the accuracy of a waveform area integration. Considering the dead time requirement, only a limited number of sampling points can be summed to obtain the results, while it takes a quite long time for the signal to drop completely to the baseline along its trailing edge. This can cause an error in the charge measurement results. Besides, there exists an uncertainty in sampling positions of a waveform, because an input signal is not correlated to the sampling clock of the ADC, which is another cause of a measurement error. To address this issue, a TDC-based correction method is proposed in this paper to enhance the precision of charge measurement further.

Through the precise measurement of a time interval difference between the input signal and sampling clock using the TDC and system calibration, the above errors could be corrected. In this paper, we analyze the relationship between the time and charge measurement results, and propose two correction algorithms to improve the charge measurement resolution without increasing the ADC sampling rate. Both simulations and tests are conducted to validate the proposed correction method, and evaluate its performance using a front-end analog electronics (FAE) module [17], which had been designed for the LHAASO WCDA.

The rest of the paper is organized as follows. In Section II, the charge measurement process in the LHAASO WCDA FAE was simulated and three main factors influencing the precision of the charge measurement were discussed. Based on the simulation results in Section II, two feasible charge measurement correction algorithms were proposed and discussed in details in Section III. In Section IV, the correction algorithms were implemented in FGPA devices, and real-time

Manuscript received Mar. 24, 2018. This work was supported in part by the Knowledge Innovation Program of the Chinese Academy of Sciences under Grant KJCX2-YW-N27 and in part by the CAS Center for Excellence in Particle Physics (CCEPP).

The authors are with the State Key Laboratory of Particle Detection and Electronics, University of Science and Technology of China, Hefei 230026; and Department of Modern Physics, University of Science and Technology of China, Hefei 230026, China (telephone: 086-0551-63607746, corresponding author: Lei Zhao, e-mail: zlei@ustc.edu.cn).

© 2018 IEEE. Accepted version for publication by IEEE. Digital Object Identifier 10.1109/TNS.2018.2878721

tests were conducted to evaluate the performance of each correction algorithms. Discussions and conclusions are presented in the last two sections, respectively.

## II. METHODS AND IMPLEMENTATIONS

### A. Charge measurement in the LHAASO WCDA FAE

The block diagram of the time and charge measurement circuit in the LHAASO WCDA FAE is presented in Fig. 1. It is based on the amplification, shaping, and waveform integration technique. The input current signal from the PMT is firstly converted to the voltage signal through $R_0$ (whose value is set to 50 Ω to achieve the impedance match) and then amplified by $A_1$. Afterwards, the signal is fed to the $RC^2$ shaping circuit with the time constant of 40 ns. The output signal from $A_3$ is digitized by the 12-bits 62.5-Msps ADC. The time measurement is based on the leading edge discrimination and the FPGA (Field Programmable Gate Array)-based TDC technique [16, 24-29]. In order to extract the charge information, the output data stream of the ADC was fed to the FPGA for peak detection or waveform integration. The signal from $A_1$ is further amplified by $A_4$ to achieve a high slew rate, then it is AC-coupled, and finally, fed to the discriminator to get the time-over-threshold information.

The FPGA-TDC implemented in the LHAASO WCDA FEA is designed based on multi-phase clock interpolation technique, which has already been finished in our previous work. A 62.5 MHz system clock is fed to the internal PLL inside the FPGA device (XC7A200T-FFG1156 in Artix-7 Series of Xilinx Inc.), which generates four synchronized 375-MHz clock signals with 0°, 45°, 90°, and 135° phases. Using these clocks combined with flip-flops within ISERDESE, a TDC bin size of 333 ps is achieved [16].

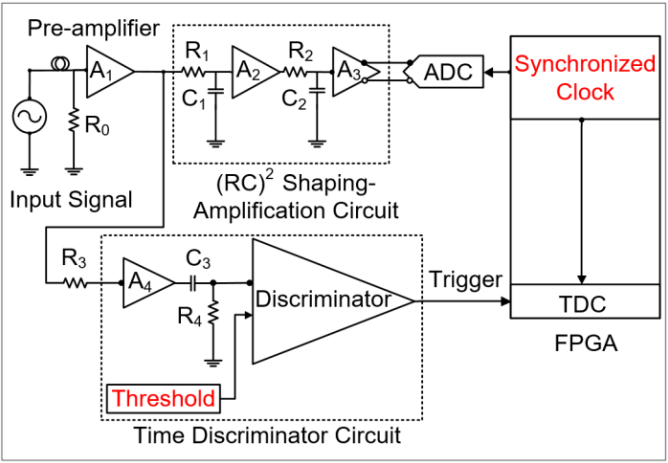

Fig. 1. The time-charge measurement FAE in the LHAASO WCDA.

We verified our proposed method based on the above WCDA readout electronics in LHAASO. First, we analyzed the charge measurement errors with simulations based on the PSpice and MATLAB software, which are presented in the following two subsections. After the discussions on our correction methods in Section III, we finally implemented the methods in the FAE (in Fig. 1) and conducted tests.

### B. Analysis of charge measurement results versus sampling uncertainty

In order to study the amplification-shaping charge measurement process in details as well as to analyze the cause of the measurement error, the PSpice simulation was conducted on the FAE (excluding ADC and FPGA components) presented in Fig. 1. In the simulation, we mimicked an input signal according to the PMT output with the amplitude of 200 mV and the leading and trailing edges of about 4 ns and 8 ns, respectively. The simulated waveforms of the input signals fed to the ADC and time discriminator are shown in Fig. 2. Then we used MATLAB to simulate the A/D conversion process to digitize the shaped waveform, also shown in Fig. 2. Wherein the black crosses refer to the sampling points of the waveform based on MATLAB simulation.

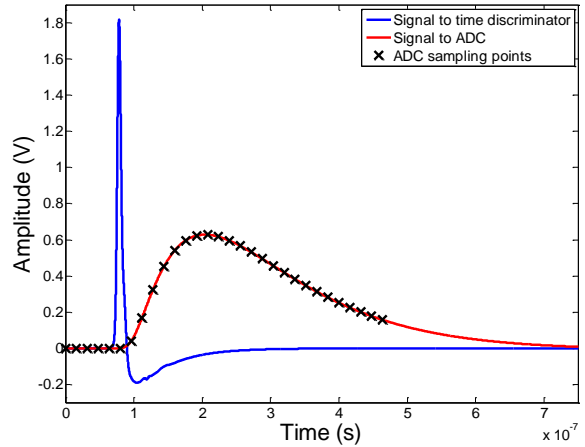

Fig. 2. Signal waveform simulation results of the charge measurement circuit.

A rough relationship between the charge measurement resolution (RMS over a mean value of a charge measurement result) of peak detection and summation is given by (1), where $N$ denotes the total number of summation points of a waveform. According to (1), by using the summation a significantly better resolution than that of the peak detection method can be achieved. That is why we chose the summation/integration method for conducting a charge measurement.

$$Resolution(summation) \sim \frac{Resolution(peak\ detection)}{\sqrt{N}}. \quad (1)$$

As mentioned above, considering the dead time of a summation, only a limited number of ADC sampling points can be added when calculating the total area of a waveform. Since the ADC sampling clock is not correlated to the input signal, a certain uncertainty exists. Therefore, we further analyze these issues in Part B and Part C.

In order to further analyze the measurement error caused by a limited number of summation points (considering the dead time) and uncertainty of a position of the first summation point, we used a TDC fine time measurement result and a quantization method to analyze the charge measurement results. As aforementioned in Part A, the TDC (bin size = 333 ps) was synchronized with the ADC sampling clock, and the period of the former one was 1/48 of that of the latter one. As it is shown in Fig. 3, when the input signal of the discriminator was equal to its threshold, the discriminator output a hit signal to the FPGA-based TDC at the time point marked as $T_{th}$. Since the ADC clock period ($T_1$ to $T_2$ in Fig. 3) was divided by the synchronized TDC clock equivalently into 48 bins, we could use the TDC fine time (marked as $T_{fine}$, which was an integer ranging from 1 to 48) to approximate $T_{th}$.

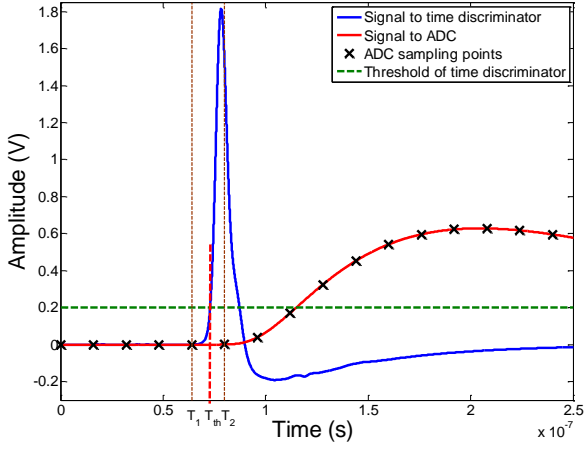

Fig. 3. The fine time of the charge measurement.

Equation (2) represents the expression of the TDC fine time ($T_{fine}$), where $Bin\_Size_{TDC}$ is the bin size of the TDC clock and [*] denotes the rounding down operation of the inner value.

$$T_{fine} = \left[\frac{T_{th}-T_1}{Bin\_Size_{TDC}}\right]+1 \qquad (2)$$
$$= \left[\frac{T_{th}-T_1}{Clk\_Period_{ADC}} \times 48\right]+1 = \left[\frac{T_{th}-T_1}{T_2-T_1} \times 48\right]+1$$

According to the FAE structure shown in Fig. 1, both charge measurement signal sampled by the ADC and input hit signal of the TDC were generated by the identical signal from the PMT. Thus, we could know the uncertainty of the ADC sampling position through the TDC fine time measurement results, which could be affected by a length difference between the waveform integral signal and timing trigger signal. Namely, the relationship between the charge measurement results and a TDC fine time provides an interesting method to utilize a TDC fine time to analyze an error of the charge measurement results and correct it.

Simulations of the ADC sampling and FPGA summation coupled with the TDC fine time were conducted using MATLAB software. The relationship between the measured charge and TDC fine time is presented in Fig. 4, while the normalized charge measurement results are presented in Fig. 4 (B), where the deviation form 1 denotes the relative error of the charge measurement.

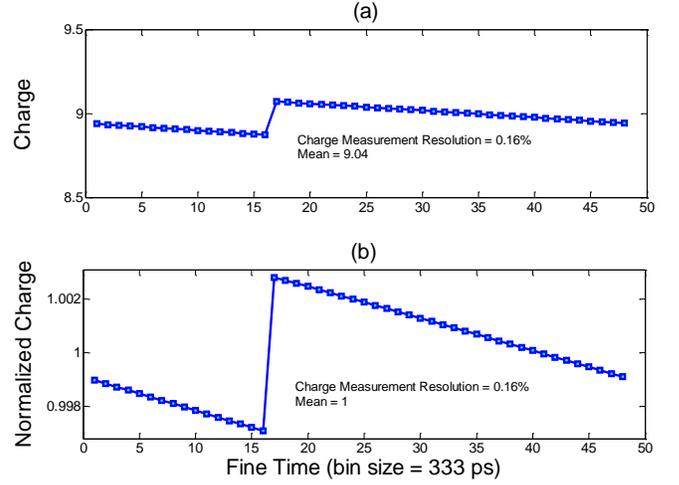

Fig. 4. The relationship between the charge measurement results and TDC fine time: (A) without normalization, (B) with normalization.

In Fig. 4, a special pattern is indicated; namely, the charge measurement result decreased when the fine time increased, except that the charge measurement result jumped from the minimum value to the maximum value at a certain fine time point called the jumping point. This phenomenon can be explained by results presented in Fig. 5. In Fig. 5(A), an identical signal located at different time positions (marked as the blue and red curves) is shown. The presented signal was fed to the time discriminator, and trigged at $T_{S1}$ and $T_{S2}$, respectively. In Fig. 5(B), the signals obtained after the shaper are shown, and they were fed to the ADC for digitization. The digitization process is based on MATLAB simulation. Although two waveforms shown in Fig. 5 (B) are the same, the positions of sampling time points on each waveform are different, so the charge measurement results also differ.

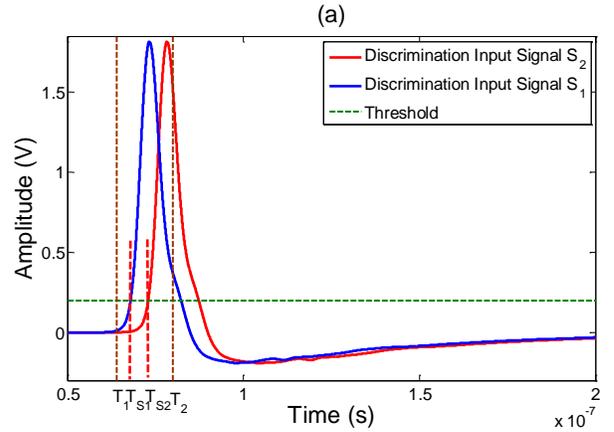

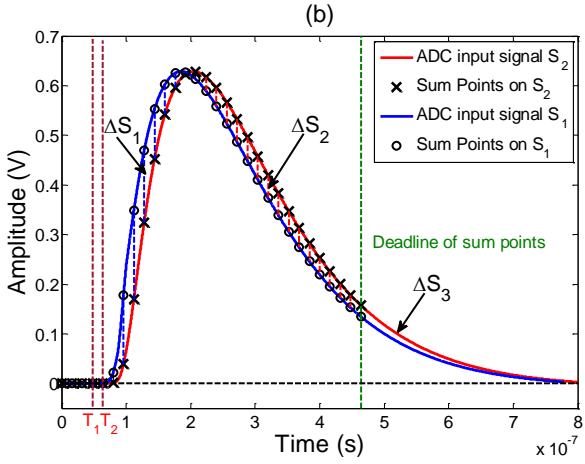

Fig. 5. (A) Input signal waveforms of the discriminator, (B) input signal waveforms of the ADC.

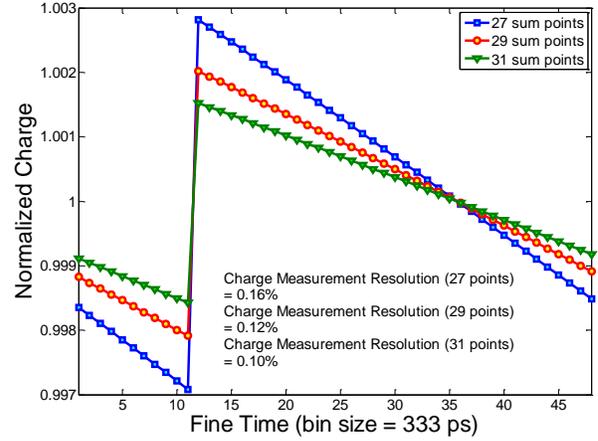

Fig. 6. The normalized charge measurement results versus the fine time at a different number of summation points.

As shown in Fig. 5(A) when the time interval between $T_{S1}$ and $T_{S2}$ is small, i.e., $T_{S1} - T_{S2} < Clk\_Period_{ADC} = 48ns$, the first point of the digitized waveform for summation is the same for signal waveforms $S_1$ and $S_2$ in Fig. 5(B), and it is obvious that the summed area of $S_1$ is larger than that of $S_2$ ( $Sum_{S1} - Sum_{S2} = \Delta S_2 - \Delta S_1 = -\Delta S_3 < 0$ ), which explains why the charge measurement result decreases with the TDC fine time. However, when ($T_{S1}$-$T_{S2}$) gradually increases until it exceeds one $Clk\_Period_{ADC}$, the first summation point of $S_2$ is delayed by 1 $Clk\_Period_{ADC}$. Thus, the total area after summation returns back to the original value, which corresponds to the jumping point in Fig. 4.

Another phenomenon is that the jumping point is neither 1 nor 48, which is analyzed in the following subsection.

## C. Further Discussions on Charge Measurement Error

Based on simulation results obtained by MATLAB software, three main influencing factors of the charge measurement precision are discussed. The discussion and simulation results are as follows.

The simulation results of the charge measurement error at different numbers of summation points are shown in Fig. 6, and they can be easily understood because it is logical that a higher precision can be achieved with a larger number of sampling points included in the integration area. Certainly, the increase in the number of summation points enlarges the dead time of a charge measurement, which is not favorable for the improvement of a charge measurement resolution.

The charge measurement error decreases with more number of sample points used. If more sample numbers are used, longer dead time will be inevitable. In this paper, we focus to correct the error with less sample points used in the summation. In the following simulation parts, the number of sampling points is set to 27.

The second factor influencing the relationship between the measured charge and the TDC fine time is the signal amplitude. The simulation results at three different signal amplitudes: 1.0 V, 1.2 V, and 2.0 V, are presented in Fig. 7(A), wherein it can be observed that the charge measurement results decrease as the fine time increases, and the decreasing trends are identical for all these three curves in Fig. 7(A) while the jumping points are different.

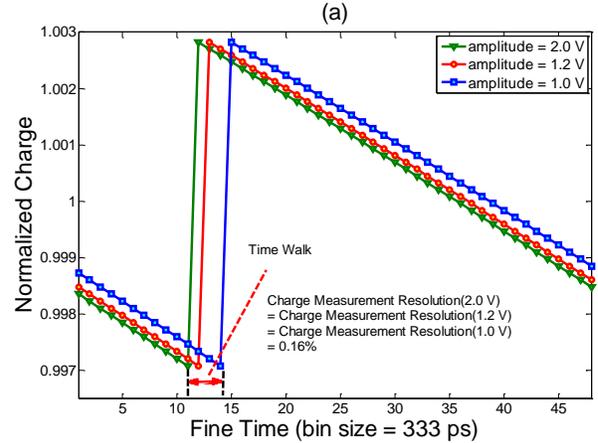

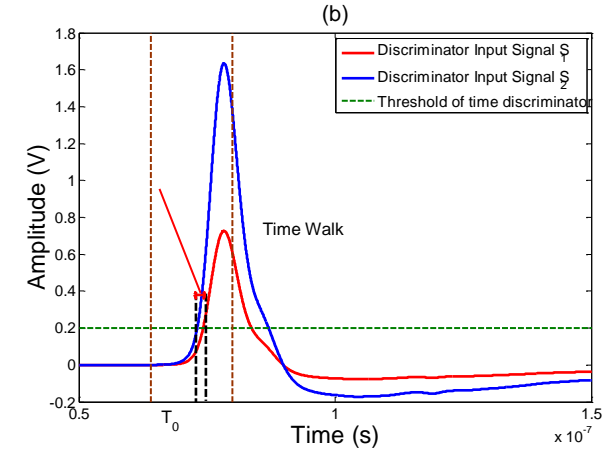

Fig. 7. (A) The normalized charge measurement results versus the fine time at different signal amplitudes, (B) time walk caused by different waveform amplitudes.

Through the analysis of the mentioned issue, we found that it is caused by a time walk shown in Fig. 7(B). At different signal amplitudes and constant threshold, the discrimination time point (corresponding to the TDC fine time) differs from $T_0$ (the real start time of the signal) as shown in Fig. 7(B), and the difference between them is larger at smaller signal amplitude, which explains the jumping point in Fig. 4 drifts at different signal amplitudes. Therefore, the calibration of the FAE circuit should be conducted, and the Look-Up Tables (LUTs) should be established for correction.

The third influencing factor is the time jitter between the TDC and the ADC clock period, as shown in Fig. 8(A). This jitter causes an error of the TDC fine time compared to the ADC clock. To study this effect, we conducted the simulations, and the results obtained by the Monte-Carlo simulation in MATLAB software are presented in Fig. 8(B). By comparing the results presented in Fig. 8(B) with those in Fig. 4, a medium transition point is observed between the two points with the minimum and maximum charge value. This is because the time jitter caused the TDC measurement error, which made some of the TDC fine times corresponding to the maximum charge results be reduced by a TDC bin (which corresponded to the minimum charge measurement, Fig. 4), and vice versa. The average effect generated the medium transition point in Fig. 8 (B). This effect furtherly shows that calibration of the relationship between the charge measurement and the TDC fine time is indispensable.

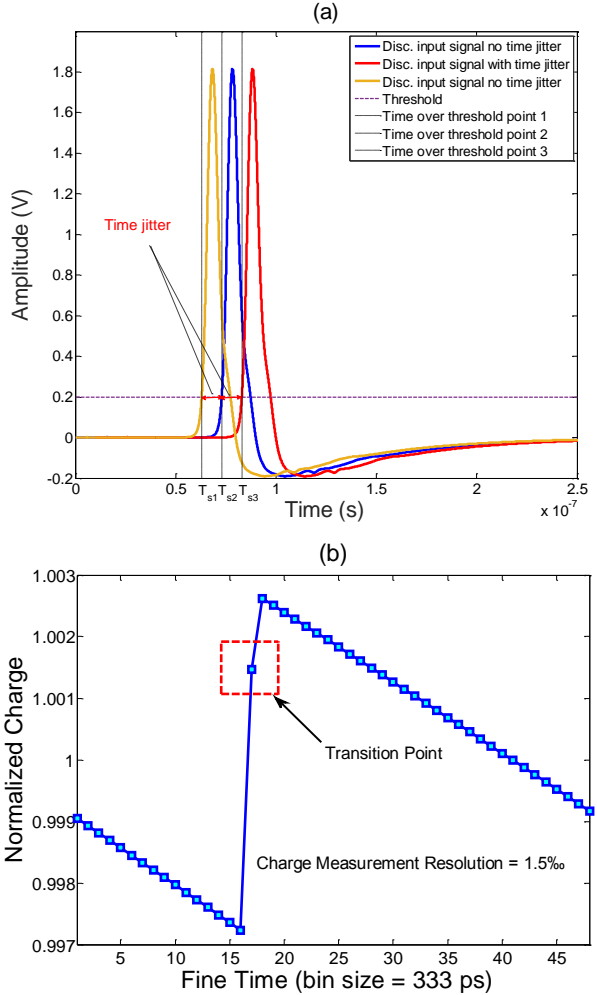

Fig. 8. (A) The time jitter of the discriminator input signal; (B) the normalized charge measurement results versus the fine time of the Monte-Carlo simulation on a time jitter.

## III. FEASIBLE CORRECTION METHODS

After obtaining the simulation results of the relationship between the charge and the TDC fine time, we propose two feasible correction algorithms which can be implemented in FPGA devices in order to improve the accuracy of the charge measurement results. The overall block diagram of the correction logic implemented in the LHAASO WCDA is represented in Fig. 9. The measured charge and TDC fine time are transferred to the correction block. Then the corrected charge measurement results and fine time will be packaged and transferred to the data interface block.

Before introducing feasible correction algorithms, we need to define the correction coefficient $C_i$:

$$C_i = \frac{\overline{Q}}{Q_i} \quad (3),$$

where $Q_i$ is the charge measurement result corresponding to the fine time $i$ ($i$ is an integer that represents the TDC bin number), and $\overline{Q}$ is the mean value of $Q_i$ ($i$ = 1, 2, …, 48). In fact, $C_i$ represents a reciprocal value of the normalized charge measurement value.

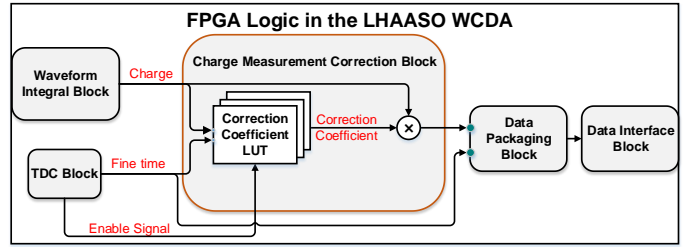

Fig. 9. The block diagram of the FPGA logic implemented in the LHAASO WCDA

### A. Dual-LUTs Correction Algorithm

The first correction algorithm is based on the dual LUTs, as shown in TABLE 1. Namely, LUT I is used for the time walk correction, which actually merges the multiple curves in Fig. 7 into one, and LUT II is used to correct the charge measurement error using a common curve (Fig. 4). In LUT I, the time walk range is divided into several zones where each zone corresponds to a charge range, e.g., $Q_{2L}$ to $Q_{2H}$, and the division in zones is performed such that in each zone the time difference variation is within one TDC bin, which means that only one fine time exists in one zone. In LUT II, there are a total of 48 cells which correspond to the ratio between the clock periods of the ADC and TDC, and the content of each cell is the correction coefficient $C_i$ given by (3).

TABLE 1
LUT I

| Range of Charge | $Q_{1L}$-$Q_{1H}$ | $Q_{2L}$-$Q_{2H}$ | $Q_{3L}$-$Q_{3H}$ | … | $Q_{mL}$-$Q_{mH}$ |
|---|---|---|---|---|---|
| Time walk | 0 | $\Delta t_1$ | $\Delta t_2$ | … | $\Delta t_m$ |

LUT II

| Fine time | 1 | 2 | 3 | … | 48 |
|---|---|---|---|---|---|
| Coefficient | $C_1$ | $C_2$ | $C_3$ | … | $C_{48}$ |

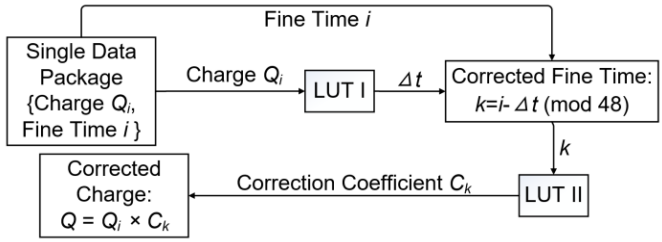

Fig. 10. The block diagram of the dual LUTs correction algorithm.

The block diagram of the dual-LUTs correction algorithm is presented in Fig. 10. The inputs data contain the charge measurement value $Q_i$ and fine time value $i$ given by (3). We firstly locate the zone in LUT I using $Q_i$ and correct the fine time to value $k = i - \Delta t$ (modulo operation, mod 48, is used when the time walk exceeds one ADC clock period). Then, we use $k$ as the index for LUT II and find the corresponding $C_k$, and the charge measurement result can be corrected to the right value by $Q = Q_i \times C_k$.

However, there is a problem due to the assumption on the LUT I that the multiple curves (Fig. 7(A)) have the same shape, while may not be the case in real applications. This means that the correction algorithm may be inconvenient for the high-resolution charge measurement applications. To address this issue, we propose another method which is presented in the following subsection.

### B. Sectional Look-Up Table Correction Algorithm

With the aim to overcome the above-explained problem of the previous correction algorithm, an alternative correction algorithm based on a sectional LUT shown in TABLE 2 is proposed. The sectional LUT divides the range of charge into several sections. The range of each section is the same as in LUT I; thus, in each section, there is no time walk. The only difference in these divisions is that here we set an individual LUT for each section instead of sharing a common LUT which is used in the previous algorithm (i.e., LUT II in TABLE 1). The block diagram of the sectional LUT correction algorithm is shown in Fig. 11. The basic idea is to locate the section where the charge measurement $Q_i$ is (e.g., the $k_{th}$ section: $Q_{kL}$-$Q_{kH}$), and then, to use the fine time $i$ to obtain the corresponding coefficient $C_{k,i}$, and finally, the correction is done by $Q = Q_i \times C_{i,k}$.

TABLE 2

| Range of Charge | Basic LUTs | | | |
|---|---|---|---|---|
| $Q_{1L}$-$Q_{1H}$ | Fine time | 1 | 2 | … | 48 |
| | Coefficient | $C_{1,1}$ | $C_{1,2}$ | … | $C_{1,48}$ |
| $Q_{2L}$-$Q_{2H}$ | Fine time | 1 | 2 | … | 48 |
| | Coefficient | $C_{2,1}$ | $C_{2,2}$ | … | $C_{2,48}$ |
| … | | | … | | |
| $Q_{mL}$-$Q_{mH}$ | Fine time | 1 | 2 | … | 48 |
| | Coefficient | $C_{m,1}$ | $C_{m,2}$ | … | $C_{m,48}$ |

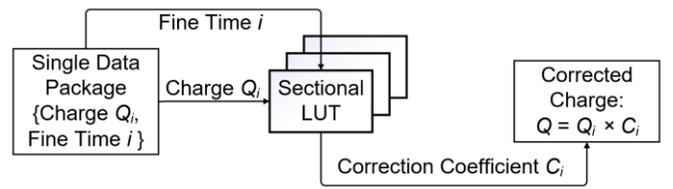

Fig. 11. The block diagram of the sectional LUT correction algorithm.

### C. Problems in Implementing Look-Up Tables

To apply the dual-LUTs correction algorithm or the sectional LUT correction algorithm, two issues have to be addressed.

The first one is to divide the range of charge into several sections in LUT I of TABLE I or TABLE 2 appropriately. Considering the inevitable time jitter in electronics, one charge result (i.e., waveform amplitude) can correspond to more than one fine time result. This phenomenon makes the division of the range of charge into appropriate small sections more difficult (statistically only one fine time exists in each section).

Our solution for this problem is to test the relative time difference between a time discrimination signal and a reference signal (e.g., a synchronized square wave trigger signal from the signal generator). Therefore, we conducted a series of experimental tests by changing the input signal amplitude and obtained a large number of combinations of time walk values and input signal amplitudes. The experimental setup for the tests is based on the LHAASO WCDA FAE, which is descried in details in Section IV. Part A. We categorized the results and plotted the histogram of test results for a certain time walk, as shown in Fig. 12. Based on our test results, a total of four time differences were found, so there are four histograms in Fig. 12, together with their Gaussian fitting curves.

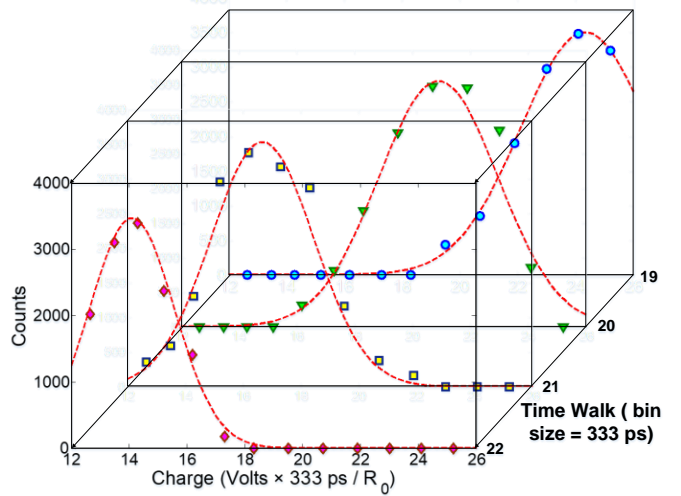

Fig. 12. Histograms of charge measurement results for different time walk values.

For the test results presented in Fig. 12, we employed the multiple Bayesian hypothesis testing [30-31] to make the division of the range of charge into appropriate small sections. We marked the hypothesis testing value for a different time walk (48 possible fine time codes) as $H_i$ ($i=1,2 … 48$). In order to optimize the division of a charge range into small sections, we minimized the statistical average decision risk by:

$$\overline{C} = \sum_{i=1}^{48}\sum_{j=1}^{48} C_{ij} P(D_i | H_j) P(H_j), \quad (4)$$

where $P(H_i)$ is the prior probability of $H_i$ (i.e., the number of events with the same time walk $i$ over the total number of events in the test), $P(D_i|H_j)$ is the probability of making decision that a time walk value equals to $i$ under the hypothesis that the time walk value is equal to $j$, and $C_{ij}$ is the cost coefficient of making such a decision.

We can rewrite (4) by:

$$\overline{C} = \sum_{i=1}^{48} P(H_i) \times C_{ii} \times \int_{x \in R_i} f(x|H_i)dx + \sum_{i=1}^{48}\sum_{\substack{j=1 \\ j \neq i}}^{48} P(H_j) \times C_{ij}$$
$$\times \int_{x \in R_i} f(x|H_j)dx \quad (5)$$
$$= \sum_{i=1}^{48} P(H_i) C_{ii} + \sum_{i=1}^{48}\sum_{\substack{j=1 \\ j \neq i}}^{48} P(H_j) \times (C_{ij} - C_{ii}) \times f(x|H_j)dx$$

where $x$ is the charge measurement value, and $f(x|H_j)$ is the probability density function of $P(D_i | H_j)$; $f(x|H_j)$ can be obtained by the Gaussian fitting of a histogram (the Gaussian fitting curve in Fig. 11).

Noticing that $\sum_{i=1}^{48} P(H_i) \times C_{ii}$ is a constant value, we define the test decision function $I_i(x)$ by:

$$I_i(x) = \sum_{\substack{j=1 \\ j \neq i}}^{48} P(H_j) \times (C_{ii} - C_{ij}) \times f(x|H_j). \quad (6)$$

The minimum decision risk can be obtained if and only if each $I_i(x)$ ($i = 1, 2, …, 48$) takes the minimum value.

To simplify the problem, we assume that:

$$C_{ij} = 1 \ (i \neq j), \ C_{ii} = 0, \quad (7)$$

which means that the cost of a false decision is 1, and the cost of a true decision is 0. Then, $I_i(x)$ can be simplified to (8). For each charge measurement value $x$, if $I_i(x) = \min\{I_1(x), I_2(x), …, I_{48}(x)\}$, and then, $x$ is discriminated into the $i^{th}$ category of time walk.

$$\Rightarrow I_i(x) = -\sum_{\substack{j=1 \\ j \neq i}}^{48} P(H_j) \times f(x|H_j). \quad (8)$$

According to the test results shown in Fig. 11, we can find that $P(H_{22}) \approx 0.23$, $P(H_{21}) \approx 0.26$, $P(H_{20}) \approx 0.27$, $P(H_i) \approx 0.24$, $P(H_{else}) = 0$, and $f(x|H_i)$ has the Gaussian function form:

$$f(x|H_i) \approx A_i \exp\left(-\frac{(x-Q_i)^2}{2\sigma_i^2}\right), \quad (9)$$

where parameters $A_i$, $Q_i$, and $\sigma_i^2$ represent the amplitude, mean, and variance of the Gaussian function, respectively.

The second problem relates to the calculation of an optimal LUT in a certain range of the charge, i.e., LUT II in TABLE 1 and each LUT for each sectional range of the charge in TABLE 2.

Our solution is as follows. Based on the aforementioned methods, we first divide the range of measured charge into several sections. Then, for each section, we select several charge values (i.e., the waveform amplitudes). The values presented in TABLE 3 correspond to one certain section, and for each amplitude $A_i$ within that section, we measure the normalized charge value $m_{i,j}$ at different fine time $j$.

TABLE 3

| Fine time / Amplitude | 1 | 2 | … | 48 |
|---|---|---|---|---|
| $A_1$ | $m_{1,1}$ | $m_{1,2}$ | … | $m_{1,48}$ |
| $A_2$ | $m_{2,1}$ | $m_{2,2}$ | … | $m_{2,48}$ |
| … | | … | | |
| $A_n$ | $m_{n,1}$ | $m_{n,2}$ | … | $m_{n,48}$ |

Then, we utilize the values presented in TABLE 3 to calculate the optimal correction coefficients (e.g., $C_1$ - $C_{48}$ in LUT II). For each $A_i$, the charge measurement resolution before and after correction can be expressed by (10) and (11), respectively:

$$F_i|_{origin} = \frac{Var\{m_{i,k}\}}{Mean\{m_{i,k}\}} = \frac{\sum_{k=1}^{48}(m_{i,k} - \overline{m_i})^2}{48 \times \overline{m_i}} \ (i=1,2\cdots,n \ \ k=1,2\cdots,48), \quad (10)$$

$$F_i|_{corrected} = \frac{Var\{C_k m_{i,k}\}}{Mean\{m_{i,k}\}} = \frac{\sum_{k=1}^{48}(C_k m_{i,k} - \overline{m_i})^2}{48 \times \overline{m_i}}. \quad (11)$$

The overall charge resolution in the charge section of TABLE 3 can be further calculated by:

$$F(C_1, C_2, \cdots C_{48}) = \sum_{i=1}^{n} F_i|_{corrected}(C_1, C_2, \cdots C_{48}) = \sum_{i=1}^{n} \frac{\sum_{k=1}^{48}(C_k m_{ik} - \overline{m_i})^2}{48 \times \overline{m_i}}. \quad (12)$$

According to the knowledge on multi-variable calculus, there exists the minimum value of $F$ given by (13):

$$\left(\frac{\partial^2 F}{\partial^2 C_1}, \frac{\partial^2 F}{\partial^2 C_2}, \cdots \frac{\partial^2 F}{\partial^2 C_{48}}\right) \geq \vec{0}. \quad (13)$$

Then, we let the first derivate of $F$ over each $C_i$ be equal to zero, and then obtain the correction coefficient $C_i$ of the optimal LUT, which is given by:

$$\left(\frac{\partial F}{\partial c_1}, \frac{\partial F}{\partial c_2}, \cdots, \frac{\partial F}{\partial c_{48}}\right) = \vec{0} \Rightarrow c_k = \frac{\sum_{i=1}^{n} m_{i,k}}{\sum_{i=1}^{n} \frac{m_{i,k}^2}{\overline{m_i}}}. \quad (14)$$

IV. INITIAL EVALUATION AND TEST RESULTS

In this section, we conducted tests to evaluate the performances of the proposed correction algorithms, namely, the dual-LUT correction algorithm and the sectional LUT correction algorithm. The correction logics of each algorithm were implemented in the FPGA devices, and the real-time performances were evaluated. The FAE module we used for the test is designed for the LHAASO WCDA. The leading and trailing edge of the signal from shaping circuits are around 80 ns and 280 ns, respectively, and the shaped signal was digitized by ADC at a sampling rate of 62.5 Msps. In order to evaluate the correction effects, we focus to correct the error with less sample points (long dead time) used in the summation. In the following parts, the number of summation points is set to 27 so that the sampling process stopped before the signal dropped completely to the baseline along its trailing edge.

## A. Test bench setup

The diagram of the experimental platform is shown in Fig. 13. We used an arbitrary signal source AFG3252 (from Tektronix Corporation) to generate the input signal similar to that of a PMT (Hamamatsu model R5912). The programmable attenuator was used for changing the amplitude of the input signal. The input pulses were fed into the LHAASO WCDA FAE, which contained the charge measurement correction logic implemented in an FPGA (Artix-7 Series of Xilinx Inc.). The LHAASO WCDA FAE communicate with our DAQ computer through a White Rabbit (WR) Switch developed by CERN.

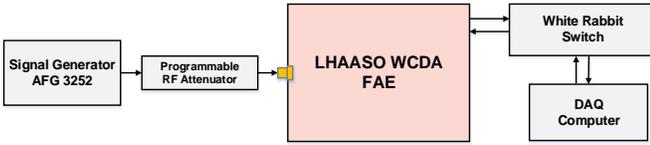

Fig. 13. The experimental platform of the correction performance evaluation.

TABLE 4 show the resource consumption of the dual LUT correction algorithms and sectional LUT correction algorithms. These results have been obtained with Vivado IDE. TABLE 4 demonstrates that both of the correction algorithms consume acceptable resources in our current application.

TABLE 4

| Correction Algorithm | | | Dual LUT | Sectional LUT |
|---|---|---|---|---|
| Resource consumption | LUT RAM | Used | 1008 | 648 |
| | | Available | 46200 | 46200 |
| | | Utilization | 2.2 % | 1.4 % |
| | LUT | Used | 12088 | 6792 |
| | | Available | 133800 | 133800 |
| | | Utilization | 0.9 % | 0.5 % |
| System clock frequency | | | 62.5 MHz | |

## B. Evaluation of Real-time Correction Performance

The normalized charge measurement result versus the fine time is presented in Fig. 14. These two curves correspond to two arbitrary signal amplitudes (1.2 V and 1.8 V). The waveforms of these two curves concord well with the predicted ones, as it can be seen in Fig. 7(A) and Fig. 8(B), which verifies the analysis given in Section II.

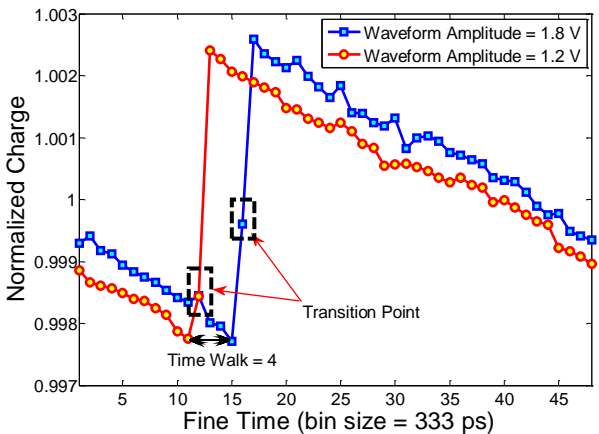

Fig. 14. The normalized charge measurement results versus fine time.

The results before and after correction are presented in Fig. 15, wherein it can be seen that the correction results of the sectional LUT at the amplitude of 2.0 V. It is similar with resulted of other situations at different amplitudes. As it can be observed in Fig. 15, the variation of the normalized charge result was significantly reduced by using the proposed correction algorithms.

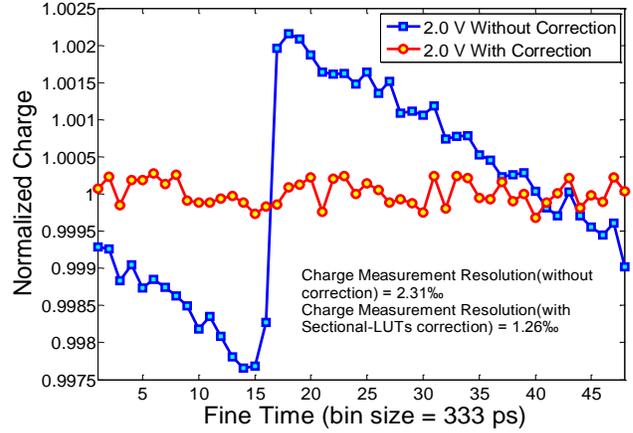

Fig. 15. The normalized charge measurement results before and after correction.

The histogram of the charge measurement results before the correction (charge resolution ~ 2.31‰) at the input signal amplitude of 2.0 V is presented in Fig. 16(A). The histograms after applying the dual-LUT method and the sectional LUT correction method are respectively presented in Fig. 16(B) and Fig. 16(C), and the charge resolution is enhanced to 1.80‰ and 1.26‰ (RMS/mean), respectively.

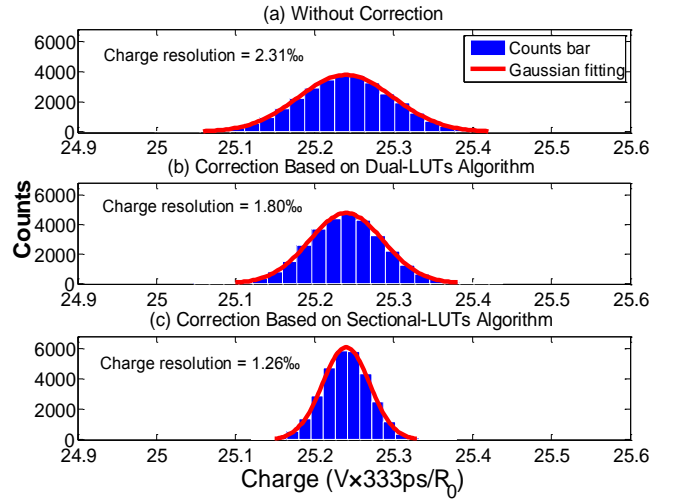

Fig. 16. The histogram of charge measurement results before and after correction.

We also conducted the tests to evaluate the correction performance over a certain range of the input signal amplitude. As shown in Fig. 17, the charge resolution was enhanced by both correction methods. Comparing the results of these two correction methods it can be found that the sectional LUT method exhibited better performance which was because of the multiple curves in Fig. 7(A) had different shapes, as aforementioned in Section III.A.

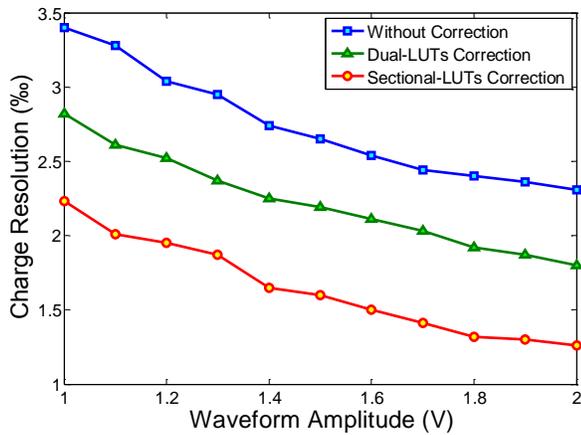

Fig. 17. The comparison of the two algorithms performance at a wide range of signal amplitudes.

## V. Discussion

According to the analysis, simulations, and test results, both dual-LUTs method and sectional LUT correction method exhibited good performance. The latter has better performance at the cost of a bit higher complexity and more FPGA resource consumption, but still acceptable in our application. Naturally, in other applications, the resource consumption could differ. And in that case, we should judge and make a tradeoff between performance and complexity, and then, decide which correction algorithm (dual-LUTs method and sectional LUT algorithm) suits better for a specific application.

## VI. Conclusion

Aiming at improving the charge measurement performance, the limiting factors of the charge measurement resolution are studied through simulation and experimental evaluation. We suggest using the TDC fine time to correct the errors in the charge measurement results. Two different correction algorithms are proposed and compared through the ral-time tests. The results indicate that the charge resolution is improved significantly by using the proposed correction algorithms, from 2.31‰ RMS to 1.26‰ RMS at input signal amplitude of 2 V.

## VII. Acknowledgement

We thank Yibao Wu from the Shenzhen Institutes of Advanced Technology, Chinese Academy of Science for his help in our research.

## References

[1] K. Bechtol et al. "TARGET: A multi-channel digitizer chip for very-high-energy gamma-ray telescopes," *Astroparticle Physics*, 2012, vol. 36, pp. 156 - 165.

[2] Stuart Kleinfelder et al. " Gigahertz Waveform Sampling and Digitization Circuit Design and Implementation," *IEEE Transactions on Nuclear Science*, 2003, vol. 50, no.4, pp. 955 - 962.

[3] S. Ritt et al. "Application of the DRS chip for fast waveform digitizing," *Nuclear Instrumentation and Methods in Physics Research A*, 2010, vol. 623, no.1, pp. 486 - 488.

[4] D. Stricker-Shaver, S. Ritt, and B. J. Pichler, "Novel calibration method for switched capacitor arrays enables time measurements with sub-picosecond resolution," *IEEE Transactions on Nuclear Science*, 2014, no. 6, vol. 61, pp. 3607 - 3617.

[5] E. Oberia et al. "A 15 GSa/s, 1.5 GHz bandwidth waveform digitizing ASIC," *Nuclear Instrumentation and Methods in Physics Research Section A*, 2014, vol. 735, pp. 452 - 461.

[6] L. Ratti et al. "A front-end channel in 65 nm CMOS for pixel detectors at the HL-LHC experiment upgrades," *IEEE Transactions on Nuclear Science*, 2017, vol. 64, pp. 789 - 799.

[7] Z. Deng et al. "Development of an eight-channel time-based readout ASIC for PET applications," *IEEE Transactions on Nuclear Science*, 2011, vol. 58, no. 6, pp. 3212 - 3218.

[8] H. Nishino, "High-speed charge-to-time converter ASIC for the Super-Kamiokande detector," *Nuclear Instruments and Methods in Physics Research A*, 2009, Vol. 10, pp. 710–717.

[9] E. Delagnes et al. "SFE16, a low noise front-end integrated circuit dedicated to the read-out of large Micromegas detectors," *IEEE Transactions on Nuclear Science*, 2000, vol. 47, pp. 1447 - 1453.

[10] D. Nitz et al. "The front-end electronics for the Pierre Auger Observatory surface array," *IEEE Transactions on Nuclear Science*, 2004, vol. 51, no. 3, pp. 413 - 419.

[11] Q. Li et al. "Front-end electronics system of PMT readout for Daya Bay reactor neutrino experiment," *Proc. IEEE NSS/MIC*, Orlando, FL, USA, 2009, pp. 1821 - 1823.

[12] X. Hao et al. "A digitalizing board for the prototype array of LHAASO WCDA," *Nuclear Science and Techniques*, 2011, vol. 22, pp. 178–184.

[13] F. P. An et al. "The detector system of the Daya Bay reactor neutrino experiment," *Nuclear Instrumentation and Methods in Physics Research Section A*, 2016, vol. 811, pp. 133 - 161.

[14] F. P. An, Q. An, J. Z. Bai et al. "Improved measurement of electron antineutrino disappearance at Daya Bay," *Chinese Physics C*, 2013, vol. 37, no. 1, pp. 011001.

[15] L. Zhao et al. "Proposal of the readout electronics for the WCDA in the LHAASO experiment," *Chinese Physics C*, 2014, vol. 38, no.1, pp. 016101.

[16] L. Zhao et al. "Prototype of the readout electronics for WCDA in LHAASO", *IEEE Transactions on Nuclear Science*, 2017, vol. 64, no. 6, pp. 1367-1373.

[17] C. Ma et al. "Analog front-end prototype electronics for the LHAASO WCDA," *Chinese Physics C*, 2016, vol. 40, no.1, pp. 016101.

[18] C. Ma et al. "Research of time discrimination circuits for PMT signal readout over large dynamic range in LHAASO WCDA," *Journal of Instrumentation*, 2016, vol. 11, pp. P11003.

[19] Z. Gu et al. "A DOI detector with crystal scatter identification capability for high sensitivity and high spatial resolution PET imaging," *IEEE Transactions on Nuclear Science*, 2015, vol. 62, no. 3, pp. 740 – 747.

[20] A. Braem et al. "Wavelength shifter strips and G-APD arrays for the read-out of the z-coordinate in axial PET modules," *Nuclear Instrumentation and Methods in Physics Research Section A*, 2008, vol. 586, pp. 300 - 308.

[21] W. W. Moses et al. "OpenPET: A flexible electronics system for radiotracer imaging," *IEEE Transactions on Nuclear Science*, 2010, vol. 57, no. 5, pp. 2532 - 2537.

[22] A. M. Williams et al, "Response of a lithium gadolinium borate scintillator in monenergistic neutron fields," *Radiation Protection Dosimetry*, 2004, vol. 110, no. 1-4, pp. 497 – 502.

[23] P. Deng et al. "Prototype design of singles processing unit for the small animal PET", *Journal of Instrumentation*, 2018, vol. 13, pp. T05007.

[24] J. Imrek et al, "FPGA based TDC using Virtex-4 ISERDES blocks", *Proc. IEEE NSS/MIC*, Knoxville, TN, USA, 2010, pp. 1413 - 1415.

[25] C. Ye et al, "A new waveform digitization based on time-interleaved A/D conversion", *Chinese Physics C*, 2013, vol. 37, no. 11, pp. 116102.

[26] C. Ye et al. "A field-programmable-gate-array based time digitizer for the time-of-flight mass spectrometry," *Review of Scientific Instrument*, 2014, vol. 85, pp. 045115.

[27] L. Zhao et al. "The design of a 16-channel 15 ps TDC implemented in a 65 nm FPGA," *IEEE Transactions on Nuclear Science*, 2013, vol. 60, no. 5, pp. 3532 - 3536.

[28] J. Wu et al. "Several Key Issues on Implementing Delay Line Based TDCs Using FPGAs," *IEEE Transactions on Nuclear Science*, 2010, vol. 57, no. 3, pp. 1543 - 1548.

[29] E. Bayer et al. "A high-resolution (<10 ps RMS) 48-channel time-to-digital converter (TDC) implemented in a field programmable gate array (FPGA)," *IEEE Transactions on Nuclear Science*, 2011, vol. 58, no. 4, pp. 1547 - 1552.

[30] J. M. Bernardo et al. "Bayesian Hypothesis Testing: A Reference Approach," *International Statistical Review*, 2002, vol. 70, pp. 351–372.

[31] S. James Press, Dec. 2002, *Subjective and Objective Bayesian Statistics:*

*Principles, Models, and Applications, Chapter 9, Second Edition*.